\documentstyle[twoside,fleqn,espcrc2]{article}

\newcommand{\AmS}{{\protect\the\textfont2
  A\kern-.1667em\lower.5ex\hbox{M}\kern-.125emS}}

\title{Singular Behaviour of the Potts Model in the Thermodynamic Limit}

\author{Ralph Kenna\address{School of Mathematics, \\
        Trinity College Dublin, \\
        Ireland}%
        \thanks{Supported bt EU TMR Project No. ERBFMBI-CT96-1757}}
       
\begin{document}

\begin{abstract}
The self--duality transformation is applied to the Fisher zeroes
near the critical point in the thermodynamic limit in the $q>4$ 
state Potts model in two dimensions. A requirement that the locus
of the duals of the zeroes be identical to the dual of the locus 
of zeroes (i) recovers the ratio of specific heat to internal energy
discontinuity at criticality and  the relationships between the 
discontinuities of higher cumulants and (ii) identifies duality 
with complex conjugation. Conjecturing that all zeroes governing 
ferromagnetic critical behaviour satisfy the latter requirement, 
the full locus of Fisher zeroes is shown to be a circle. This locus, 
together with the density of zeroes is shown to be sufficient to 
recover the singular form of all thermodynamic functions in the 
thermodynamic limit.
\end{abstract}

\maketitle

\section{INTRODUCTION}

The question of the locus of Fisher zeroes in the $d=2$ Potts 
model is one which has recently received an increased amount
of attention \cite{Martin,LeeI,AnMa94,MaSh96,ChHu96,WuRo96}.
In the two--state (Ising) case where the exact solution is 
available \cite{Ka49}, the Fisher zeroes form two circles in 
the complex $u=\exp{(-\beta)}$ plane in the thermodynamic limit 
\cite{Fi64}.

The partition function for the $q$--state Potts model is 
$ Z_L(\beta)  =  \sum_{\{\sigma_i\}} 
\exp{(\beta \sum_{\langle ij \rangle}{\delta_{\sigma_i \sigma_j}})}$
and in the thermodynamic limit is invariant under the duality 
transformation $u \rightarrow {\cal{D}}(u) $, where
$ {\cal{D}}(u) = (1-u)/(1+(q-1)u)$ \cite{Po52}.
The critical temperature at which the phase transition occurs is
invariant under duality and is $  u_c = 1/(1 + \sqrt{q})$.
Here the $d=2$ $q>4$ state model exhibits a first order phase 
transition \cite{Ba}.

Based on similarities with the Ising case, Martin and Maillard and 
Rammal \cite{Martin} conjectured that the locus of Fisher zeroes in 
the $d=2$, $q$--state Potts model be given by an extension of 
critical duality to the complex plane, namely ${\cal{D}}(u)=u^*$ 
where $u^*$ is the complex conjugate of $u$. This identification 
yields a circle with centre $-1/(q-1)$ and radius $\sqrt{q}/(q-1)$.
When $q=2$ this recovers the ferromagnetic Fisher circle of the 
Ising model \cite{Fi64}. In the Ising case, the partition function
is actually a function of $u^2$. There, the second (antiferromagnetic)
Fisher circle comes from the map $\beta \rightarrow - \beta$.
Numerical investigations for small lattices \cite{Martin} provided 
evidence that the Fisher zeroes indeed lie on the circle given by  
the identification of duality with complex conjugation. However the 
numerics are highly sensitive to the boundary conditions used and the 
situation far from criticality remained unclear. Some progress was 
recently made in the non--critical region using low temperature 
expansions for $3\le q \le 8$ \cite{MaSh96}.

Recently, and on the basis of numerical results on small lattices with
$q \le 10$, it has again been conjectured that for finite lattices with 
self--dual boundary conditions, and for other boundary conditions in 
the thermodynamic limit, the zeroes in the ferromagnetic regime are 
on the above circle \cite{ChHu96}. The conjecture of \cite{ChHu96} 
was, in fact, proven for infinite $q$ in \cite{WuRo96}. This 
circle--conjecture is similar to another recent conjecture 
\cite{AnMa94}, namely that the Fisher zeroes for the $q$--state 
Potts model on a triangular lattice with pure three--site interaction 
in the thermodynamic limit (which is also self--dual \cite{Ba}) lie 
on a circle and a segment of the negative real axis. 

All of the above conjectures regarding the locus of Fisher zeroes
rely, at least in part, on numerical approaches. In this paper, the 
problem is addressed analytically and the origin of the circular locus 
is clarified.

\section{Thermodynamic Functions}

For finite $L$, the partition function is a polynomial in $u$ and can
be expressed in terms of its  zeroes as  
$ Z_L(\beta) \propto \prod_{j=1}^{dV}{(u-u_j(L))}$ \cite{Fi64}.
The free energy is $ \beta f(\beta) = - \ln{Z(\beta)}/V$.
The internal energy is (up to a constant)
\begin{equation}
 e(\beta)= 
 \frac{u}{V}\sum_{j=1}^{dV}\frac{1}{u-u_j(L)}
\quad .
\label{e}
\end{equation}
The general $n^{\rm{th}}$ cumulant is defined as
$ \gamma_n(\beta) = (-)^{n+1}
 \partial^n (\beta f) / \partial \beta^n $.
In conventional notation the specific heat is 
$ c(\beta) =  k_B \beta^2 \gamma_2(\beta)$.
With
$ \Delta \gamma_n  \equiv  \gamma_n(\beta^-) -
 \gamma_n(\beta^+)$, the discontinuity in the $n^{\rm{th}}$ cumulant 
at the critical temperature, the exact thermodynamic limit results 
include \cite{Ba,Wu,BhLa97,JaKa97}
\begin{equation}
  \Delta c
  = 
   k_B \beta_c^2 \frac{\Delta e}{\sqrt{q}}
\quad .
\label{diffc}
\end{equation}

\section{Partition Function Zeroes}

Recently, Lee \cite{LeeI} has derived a general theorem for first
order phase transitions in which the zeroes can be  expressed in 
terms of the discontinuities in the thermodynamic functions. For a 
system with a temperature--driven phase transition, Lee's result is
(in terms of $t=1-\beta/\beta_c$)
\begin{eqnarray}
 \beta_c {\rm{Re}} t_j(L)
 & = &
 A_1 I_j^2 + A_3 I_j^4  + A_5  I_j^6
 + \dots
\quad ,
\nonumber \\
 \pm \beta_c {\rm{Im}}t_j(L) 
 & = &
 I_j
 +
 A_2 I_j^3
 +
 A_4 I_j^5
 + \dots
\quad ,
\label{Itj}
\end{eqnarray}
where $ I_j = (2j-1)\pi/(V\Delta e)$ and  $\dots$ includes terms 
which vanish in the infinite volume limit. The first few 
coefficients $A_n$ are \cite{LeeI,Ke97}
$ A_1  =   \Delta c / 2 k_B \beta_c^2  \Delta e $,
$ A_2  =  - 2A_1^2 + \Delta \gamma_3 / 3!\Delta e$.

\subsection{The Locus of Zeroes}
\label{3.1}

From (\ref{Itj}) the real part of the zeroes (in the thermodynamic 
limit) can be expressed in terms of their imaginary parts as
$ \beta_c {\rm{Re}}t = {\cal{L}}(\beta_c {\rm{Im}}t) $
where $ {\cal{L}}(\theta) =  A_1\theta^2 + (-2A_1A_2+A_3)\theta^4
 + \dots $. The zeroes are thus seen to lie on a curve. In the complex 
$u$ upper half--plane the equation of this curve is 
$ \gamma^{(+)}(\theta) = u_c \exp{({\cal{L}}(\theta)+i\theta)}$.

\subsection{The Dual of the Locus and the Locus of the Duals}

Applying the duality transformation to
$\gamma^{(+)}(\theta)$ and expanding in $\theta$ gives
$ {\rm{Re}}  {\cal{D}} \left(   {\gamma^{(+)}}(\theta) \right)
 = u_c [  1 +  (\theta^2/2q)(2\sqrt{q}-2A_1q -q )   + \dots ] $ and
${\rm{Im}}  {\cal{D}}\left(   {\gamma^{(+)}}(\theta) \right) = 
-u_c [\theta -(\theta^3/6q)(6-6q^{1/2}+q-12A_1q^{1/2}+6A_1q) + \dots]$.

Alternatively, applying the duality transformation directly
to the $j^{\rm{th}}$ zero  in the finite--size system gives
\begin{eqnarray}
 \beta_c {\rm{Re}} t_j^D(L)
 & = &
 A_1^D
 I_j^2
 + A_3^D
 I_j^4
 + A_5^D
 I_j^6
 + \dots
\nonumber
\\
 \mp \beta_c {\rm{Im}}t_j^D (L)
 & = &
 I_j
 +
 A_2^D 
 I_j^3
 +
 A_4^D
 I_j^5
 + \dots
\label{ItjD}
\end{eqnarray}
where terms vanishing in the thermodynamic limit are included in 
$\dots$ and the first few  $A_n^D$ are
\begin{eqnarray}
 A_1^D
 & = & 
 q^{-\frac{1}{2}}-A_1
\quad ,
\label{ad1}
\\
 A_2^D  & =  &
 -q^{-1}+2 q^{-\frac{1}{2}}A_1 + A_2
\quad .
\label{ad2}
\end{eqnarray}
As in Sec.(\ref{3.1}), (\ref{ItjD}) gives the locus of the dual of 
the upper half--plane zeroes in the thermodynamic limit to be
${\gamma^{(+)}}^D(\theta) = u_c \exp{({{\cal{L}}^D}(\theta)-i\theta)}$,
where $ {\cal{L}}^D(\theta) =  A_1^D\theta^2 +
 (-2A_1^DA_2^D+A_3^D)\theta^4 + \dots $.
The expansion of this locus of duals is
${\rm{Re}}{\gamma^{(+)}}^D(\theta)  =  u_c [1+(\theta^2/2!)(-1+2A_1^D) 
  + \dots     ]$ and
$ {\rm{Im}}{\gamma^{(+)}}^D(\theta)  =  -u_c [\theta 
 + (\theta^3/3!)(-1+6A_1^D ) + \dots ]$.

Even when the finite--$L$ system does not have duality--preserving
boundary conditions, taking the thermodynamic limit restores 
self--duality. There the dual of the locus of zeroes and the 
locus of duals must be identical,
\begin{equation}
  {\cal{D}}\left( \gamma^{(+)} \right)
  \equiv
  {\gamma^{(+)}}^D
\quad .
\label{demand}
\end{equation}
Up to ${\cal{O}}(\theta^2)$ this is trivial. To ${\cal{O}}(\theta^3)$
and (separately at) ${\cal{O}}(\theta^4)$ they are identical if
$ A_1=1/(2\sqrt{q})$. This is the result (\ref{diffc}).
The identity (\ref{demand}) at ${\cal{O}}(\theta^5)$ and (separately at)
${\cal{O}}(\theta^6)$ gives
$A_3 = A_2/\sqrt{q} - q^{-3/2}(q-3)/24$, or \cite{Wu,BhLa97,JaKa97}
\begin{equation}
 \Delta \gamma_4 = \frac{6}{\sqrt{q}}\Delta \gamma_3
 + \frac{q-6}{q^{3/2}}\Delta e 
\quad .
\label{a4}
\end{equation}
Higher order results are obtainable using a computer algebra system
such as Maple \cite{Ke97}. 

\section{The Full Locus and
the Singular Parts of the Thermodynamic Functions}

Putting the above equations ((\ref{diffc}) and (\ref{a4})) into
(\ref{ad1}), (\ref{ad2}) (and their higher order equivalents)
yields $A_j^D  =  A_j$ (this has been verified up to $j=8$).
Therefore (at least up to $\theta^{10}$) the dual of the locus of zeroes
is the complex conjugate of the original locus of zeroes. We now 
assume that this is the case for all $\theta$. Then, the full feromagnetic
locus of zeroes (that part of the full locus which intersects the real 
temperature axis at the physical ferromagnetic critical point) is found 
by identifying \cite{Martin}
${\cal{D}}\left({\gamma^{(+)}}(\theta)\right)={\gamma^{(+)}}^*(\theta)$,
where  ${\gamma^{(+)}}^*$ represents the complex conjugate of
$\gamma^{(+)}$. The full ferromagnetic locus is then \cite{Martin}
\begin{equation}
 \gamma(\theta)=\frac{1}{q-1}\left(-1+\sqrt{q}e^{i\theta}\right)
\quad .
\label{ff}
\end{equation}

The density of zeroes is given by \cite{LeeI}
\begin{eqnarray}
 \lefteqn{
2 \pi  g(\theta)= 
 \left(1+\frac{1}{(q-1)\gamma(\theta)}\right) \Big[ 1+
} 
\nonumber \\
 &  & 
    \sum_{n=2}^{\infty}{
               \frac{\Delta\gamma_n}{(n-1)!}
         \left(
        \ln{ \left(
           (\sqrt{q}+1)\gamma(\theta)
             \right)
           }
         \right)^{(n-1)}
         }
 \big] .
\label{density}
\end{eqnarray}
The internal energy is (from (\ref{e}) or \cite{Abe})
\begin{equation}
 e = {\rm{cnst.}}
     +
     u
     \int_0^{2\pi}{
                  \frac{g(\theta)}{u-\gamma(\theta)}d\theta
                  }
\quad .
\label{egl}
\end{equation}
Therefore, from (\ref{ff}), (\ref{density}) and (\ref{egl}), 
the internal energy is 
$ e(\beta < \beta_c) =  e_0 $ and
\begin{equation}
 e(\beta > \beta_c) 
  =
 e_0 
 -
 \Delta e
          + \sum_{n=2}^{\infty}
            \frac{\Delta \gamma_n  (\beta_c - \beta)^{(n-1)}
     }{(n-1)!}
,
\label{e2}
\end{equation}
with $e_0$ a constant (one expects that when separate Fisher loci 
which don't cross the positive real temperature axis are accounted 
for, $e_0$ becomes temperature dependent). At $\beta_c$ the internal 
energy discontinuity 
$e(\beta=\beta_c^-)- e(\beta=\beta_c^+) = \Delta e$ is recovered.
Appropriate differentiation recovers the discontinuities in specific
heat and higher cumulants.

\section{Conclusions}

The requirement that the dual of the locus of zeroes be identical to 
the locus of the duals of zeroes (i) recovers the ratio of specific 
heat to internal energy discontinuity at criticality and relations 
between the discontinuities of higher cumulants and (ii) identifies 
duality with complex conjugation. 

Conjecturing that all zeroes governing ferromagnetic critical 
behaviour satisfy (ii) gives that this locus is the circle (\ref{ff}).
This puts the conjecturs of \cite{Martin,AnMa94,ChHu96}
on an analytic footing. The locus (\ref{ff}), together with 
the density of zeroes (\ref{density}) is sufficient to recover the 
{\em{singular}} parts of all thermodynamic functions in the 
thermodynamic limit.
It is to be expected that the {\em{regular}} parts come from 
separate loci of zeroes which don't cross the positive
real temperature axis. 

The author thanks A. Irving, W. Janke and J. Sexton.

\end{document}